\documentclass[preprint,aps,showpacs]{revtex4}
\usepackage{graphicx}
\usepackage{mathptmx} % postscript fonts
\usepackage{natbib}

\begin{document}
%\hfill BNL-xxx       % for bnl number

\title{
Universal Earthquake-Occurrence Jumps,
Correlations with Time,
and Anomalous Diffusion
}
\author
{
\'Alvaro Corral
}
\affiliation{
%Grup de F\'\i sica Estad\'\i stica,
Departament de F\'\i sica, Facultat de Ci\`encies, 
Universitat Aut\`onoma de Barcelona,
%Edifici Cc, 
E-08193 Bellaterra, Barcelona, Spain 
}
\email{alvaro.corral@uab.es}
\date{\today}

\begin{abstract}
{
Spatiotemporal properties of seismicity
are investigated for a worldwide (WW) catalog and
for Southern California in the stationary case (SC),
showing a nearly universal scaling behavior.
Distributions of distances between consecutive 
earthquakes (jumps) are magnitude independent
and show two power-law regimes, separated by 
jump values about 200 km (WW) and 15 km (SC).
Distributions of waiting times conditioned to the value
of jumps show that both variables are correlated in general,
but turn out to be independent when only short or long
jumps are considered.
Finally, diffusion profiles reflect
the shape of the jump distribution.
}
\end{abstract}

\maketitle

Earthquake statistics provides important insights
about seismic processes \cite{Stein.2003}. 
Simple laws of seismicity
such as the Gutenberg-Richter frequency-magnitude relation 
\cite{Kagan_physD,Turcotte_book,Utsu_handbook} 
and the Omori law for the temporal decay 
of the seismic activity \cite {Utsu_handbook} 
allow for instance the prediction 
of the size of a characteristic earthquake,
the evaluation of the spatial distribution of stress in a fault,
or the estimation of seismic risk after a strong event 
\cite{predict_b_nature,Schorlemmer,Reasenberg}.
%%%OJO!!!! SON MULTIDIM????
It is natural to expect that more sophisticated,
multidimensional statistical studies can be
very valuable for hazard assessment as well as for understanding
fundamental properties of earthquakes.
In this way, the structure of earthquake occurrence
in magnitude, space, and time should reflect all the complex
interactions in the lithosphere.

The first systematic analysis of seismicity taking into account
its multidimensional nature was performed by
Bak {\it et al.}, 
who studied the dependence of waiting-time distributions
on the magnitude range and the size of the spatial region 
selected for analysis, 
Their study revealed, by means of a scaling law,  
the complexity of a self-similar dynamical process
with a hierarchy of structures over a wide range of scales
\cite{Bak.2002,Corral_pre.2003,Corral_physA.2004,Corral_Christensen}.
Waiting times (also referred to as return times, inter-event times, etc.) 
are just the time intervals between consecutive earthquakes in a 
region, and can be studied in two ways: 
i) by using single-region waiting-time distributions, 
in which an arbitrary region is characterized by its own
distribution \cite{Corral_prl.2004},
or ii) the original approach \cite{Bak.2002}, 
in which a large region is divided into smaller, equally sized areas
and waiting times are measured for the smaller areas
but are included together into a unique mixed distribution 
for the whole region. 
(This procedure can be interpreted as the measurement of
first return times of seismic activity %%%to small areas
to different points
in a large, spatially heterogeneous system,
see Ref. \cite{Corral_pre.2003}.)
The outcomes of both approaches are clearly different, but
in any case scaling turns out to be a fundamental tool of analysis,
reducing the multidimensional dependence of the waiting-time
distributions to simple univariate functions.
Further,
it is possible to understand this scaling approach in terms of
a renormalization-group transformation \cite{Corral_prl.2005}.

Equally important for risk estimations and forecasting,
though much less studied \cite{Ito,Abe}, %than waiting-time distributions
should be the statistics of the distances between consecutive earthquakes,
which we can identify with jumps (or flights) of earthquake occurrence.
Very recently, Davidsen and Paczuski have provided a coherent picture
using Bak {\it et al.}'s mixed-distributions procedure \cite{Davidsen_distance};
in contrast, our paper undertakes the study of the earthquake-distance problem
considering the simpler approach of single-region distributions,
and for the case of stationary seismicity.
The results will lead us to examine the distribution of waiting times
conditioned to different values of the jumps;
finally, we measure diffusion profiles for seismic occurrence,
turning out to be that the profiles are largely determined by
the distribution of jumps.
%%%rather than by the waiting-time distributions.

Consider 
that an arbitrary spatial region and a range of magnitudes have been
selected for analysis.
The unit vector locating the 
epicenter of the $i-$th earthquake on the portion of the Earth surface
selected is given by
$$
  \hat r_i = \left( \cos\varphi_i\cos\theta_i, \,
  \sin\varphi_i\cos\theta_i, \, \sin\theta_i\right),
$$
where the angles $\varphi_i$ and $\theta_i$ denote
longitude and latitude, respectively.
The spatial distance, or jump, between the $i-$th event and the immediately previous
in time event, $i-1$, can be obtained from the angle $\alpha_i$ defined by 
the two vectors,
$$
\alpha_i = \arccos (\hat r_{i-1} \cdot \hat r_i).
$$
In this way one can measure distances as angles, in degrees;
the distance in km is obtained by multiplying 
$\alpha_i$ in radians by the Earth radius 
(about 6370 km, assuming a spherical Earth).

Given a set of values of the jumps, their probability
density $D(\alpha)$ is defined as the probability per unit distance
that the distance is in a small interval containing $\alpha$.
%%%, i.e.,
%%%$$
%%%D(\alpha) \equiv \frac{\mbox{Prob}
%%%[\alpha \le \mbox{ distance }< \alpha+d\alpha]}{d\alpha},
%%%$$
%%%where $d\alpha$ is a small distance range, 
%%%relative to the value of $\alpha$.
In order to avoid boundary problems it is convenient to 
start the analysis of seismicity on a global scale,
which has also the advantage of stationarity
(or, more properly, homogeneity in time, at least for the last
30 years). 
Stationarity means that any short time period is described by 
(roughly) the same
seismic rate, $R$ (defined as the number of earthquakes per unit time);
in such a case, a linear increase of the cumulative number of earthquakes
versus time must be observed. 
Notice that stationarity does not mean that aftershock sequences
are not present in the data, rather, many sequences can be interwinned
but without a predominant one in the spatial scale of observation.

\begin{figure*}
\centering
\includegraphics[width=3.5in]{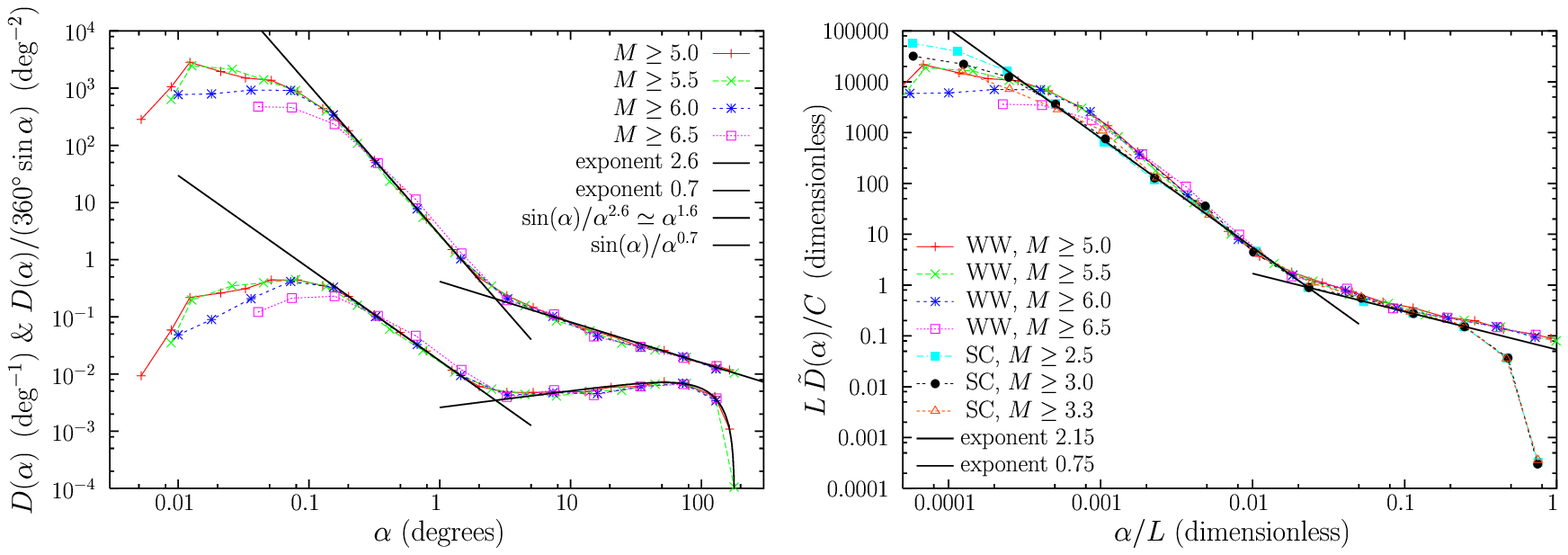}
\caption{
(color online)
(a)
Probability densities of earthquake jumps, 
defined per unit distance (angle), 
$D(\alpha)$, and per unit solid angle, 
$\tilde D(\alpha) = D(\alpha) /(360^\circ \sin\alpha)$
(shifted up a factor 1000 for clarity sake), 
for worldwide seismicity
%%%from 1973 to 2002 
with lower magnitudes
from $M_c=5$ to $M_c=6.5$.
Lines come from power-law fits to $\tilde D(\alpha)$ 
for $M_c=5$.
%%%%%%%%%%%%%%%%%%%%%%%%%%%%%%%%%%%%%%%%%%%%%%%%%%%
(b)
Rescaled probability densities 
%%%of earthquake jumps, defined per unit solid angle, 
$\tilde D(\alpha)$
for worldwide seismicity (WW) and for 
several stationary periods in Southern California (SC),
with different $M_c$ values.
For each case, $L=180^\circ$ and $L=6.5^\circ$,
and the normalization factors $C=0.0235$ and $C=1.14$ degrees$^{-1}$
are determined for $M_c=5$ and $M_c=2.5$
disregarding the smallest jumps.
Power-law fits for SC with $M_c=2.5$ are shown.
\label{L360}
}
\end{figure*}

The results for the NEIC-PDE worldwide catalog
\cite{PDE},
covering the period 1973-2002, 
are shown in Fig. \ref{L360}(a) (bottom set of curves),
using events with magnitude $M$ above different thresholds $M_c$
(i.e., $M \ge M_c$).
An important conclusion drawn from the figure
is the independence of the jump distributions on the magnitude threshold,
as in Ref. \cite{Davidsen_distance},
implying that the spatial occurrence of large earthquakes is no different
than the occurrence of small ones, in contrast to claims by some
authors \cite{Molchan04}.
There is, nevertheless, an exception 
for small distances, less than
about $0.1^\circ$ ($\simeq 10$ km) for $M_c$ from 5.5 to 6, 
comparable with the size of rupture \cite{Davidsen_Grassberger}.
%%%The discrepancies for very small $\alpha$ are due to 
%%%poor statistics with increasing $M_c$.
The distributions are characterized by
a decreasing power-law regime from about 
$0.1^\circ$ to $2^\circ$, with an exponent about 1.6,
and a possible increasing second power law for $\alpha > 2^\circ$,
with exponent 0.3, decaying abruptly close to the maximum distance. 
Note the difference between these exponents and the value 0.6, obtained
in Ref. \cite{Davidsen_distance} mixing many areas in Southern California 
and including non-stationary periods.

However, the measurement of $D(\alpha)$
contains an inconsistency when it is used for distances
over a sphere; for instance, after any event, there are
many ways to obtain a jump of $\alpha=90^\circ$ (360 ``ways'' 
in intervals of $1^\circ$) but there is only one way to 
get $\alpha=180^\circ$ (which is the maximum possible distance,
corresponding to the antipodes of the initiating event). 
Therefore, instead of working with the probability density 
defined per unit distance, i.e., per unit angle, it is preferable
to use (and easier to interpret)
the probability density defined per unit solid angle
(taking into account all the possible orientations for
a given distance $\alpha$).
A straightforward way to estimate this quantity is by means
of the transformation
$$
D(\alpha) \rightarrow \tilde D (\alpha) \equiv
{D(\alpha)}/{(2 \pi \sin \alpha)},
$$
due to the fact that $2 \pi \sin \alpha \, d\alpha$
is the element of solid angle defined by the points at distance
$\alpha$.

The results of the transformation are clearly seen in Fig. \ref{L360}(a)
(top curves). In addition to the power law for the range 
$0.1^\circ - 2^\circ$, whose exponent turns out to be $\sim 2.6$,
the second power law, now decreasing,  becomes more apparent
from about $2^\circ$ to the maximum distance,
$180^\circ$, with an exponent $\sim 0.7$.
This second power law 
seems to imply a dependence of events at large spatial scales.
Indeed, for a given direction, it is more likely a jump with $\alpha=20^\circ$
than one of $160^\circ$
(independence would imply that the probabilities were the same);
however, this effect is due to the fact that earthquake occurrence 
is not uniform over the Earth but fractal,
and the power law reflects the fractal structure of the epicenters.
If for long distances there is no causal relation between events,
the distribution of earthquake jumps is equivalent to the distribution 
of distances between any pair of earthquakes (not necessarily consecutive);
and as the density corresponding to this distance (per unit distance)
is the derivative of the well-known correlation integral, 
widely used to obtain fractal dimensions, 
independence at long distances implies
$$
D(\alpha) \sim \alpha^{d_f-1} 
\, \mbox{ and }
\tilde D(\alpha) \sim 1/\alpha^{2-d_f} 
\,\mbox{ for long $\alpha$,}
$$
where $d_f$ is the correlation dimension.
We have measured the probability density of the distance
between any two earthquakes (per unit solid angle), obtaining 
a behavior proportional to $1/\alpha^{0.85}$,
for events with $M\ge 6$, 
which implies a correlation dimension $d_f \simeq 1.15$
%%%%CITA KAGAN !!!!!
in reasonable agreement with $\tilde D(\alpha) \sim 1/\alpha^{0.7}$,
%%$\tilde D(\alpha) = D(\alpha)/\sin \alpha \sim 1/\alpha^{0.8}$,
and confirming the independence of events for $\alpha > 2^\circ$.

Turning back to the short-jump regime, 
the excess of probability given by the corresponding power-law
with respect the long-jump power law (associated to the
uncorrelated regime) is a clear sign of spatial clustering
in earthquake occurrence, extending up to distances 
of about $2^\circ \simeq 200$ km.
It is likely that this clustering extends beyond this limit, 
but it is not detectable with this procedure as it
is hidden below the uncorrelated, long-distance regime.
In any case, the power law behavior implies the no 
existence of a finite correlation length, at least up to 
200 km, at variance with the findings of Ref. 
\cite{Huc_Main,McKernon}.
In consequence we can consider the earthquake jumps
as L\'evy flights, which are one of the main causes
of anomalous diffusion.

In contrast to worldwide seismicity, regional seismicity 
turns out to be nonstationary, in general,
and in the same way as for waiting-time distributions 
\cite{Corral_prl.2004}, 
the distributions of earthquake jumps
depend on the time window selected for analysis.
This problem will be avoided here by considering specific time windows 
characterized by stationary seismicity;
an important realization then is that stationary seismicity is characterized by
stationary distributions.
Additionally, stationary seismicity has the advantage 
of a lower magnitude threshold of completeness, 
due to the absence of large events.
%%aftershock sequences.

We consider Southern-California seismicity,
from the waveform cross-correlation catalog 
by Shearer {\it et al.} \cite{Shearer}.
The analysis has been performed on 11 stationary periods,
comprised between the years 1986 and 2002,
yielding a total time span of 9.25 years 
and containing 6072 events for $M\ge 2.5$.
It is very striking, as Fig. \ref{L360}(b)
shows, that the distributions of jumps
resemble very much those of the worldwide case, 
with two power-law regimes, but in a different scale
(the separation is at about $0.1^\circ$). 
In principle, for a smaller area such as California one would
expect that the distribution of jumps is just a truncation
of the global one, but Fig. \ref{L360}(b)
clearly refutes this fact,
providing a clear illustration of earthquake-occurrence
self-similarity in space.
In fact, the figure shows these distributions 
rescaled by a factor $L$,
with $L$ the maximum distance for 
the region, which is 
$L\simeq 6.5^\circ$ for Southern California.
and $L=180^\circ$ for the worldwide case
(also included in the plot).
The reasonable data collapse is a signature of the existence
of a scaling law for the jump distribution,
$$
%   D(\alpha) \simeq \frac 1 L g\left(\frac \alpha L\right)
    D(\alpha) \simeq  g(\alpha/ L)/L
$$
(for $\tilde D(\alpha)$ we need to include an extra factor $C$ of
normalization).
The scaling law can only be given the status of approximated, 
as the exponent for short jumps for Southern California 
seems to be different than in the worldwide case,
2.15 versus 2.6 for the distribution $\tilde D$.
Nevertheless, this variation could be due to artifacts
in the short-distance properties of the catalogs.
On the other hand, the long-jump regime is well described by
the worldwide exponent.
We recall that the exponents are far from the 
value of Davidsen and Paczuski for the nonstationary California case
\cite{Davidsen_distance}.
The main difference with the worldwide case is the abrupt 
decay of $\tilde D(\alpha)$ close to the maximum $\alpha$.
Clearly, this is a boundary effect. We have tried several 
corrections for this effect (in particular the factor
$2 \pi \sin\alpha$ is not right when the ring defined by
the set of points at a distance $\alpha$ is not 
totally included in the region under study)
but we have not found a substantial improvement and
then we have preferred to keep the things simple.

The distributions of jumps we have obtained, 
together with previous work on the distribution
of waiting times \cite{Corral_prl.2004}, 
provide a simple picture of earthquake
occurrence as a continuous-time random walk.
This means that we can understand seismic activity as a
random walk over the Earth surface,
where one event comes after another 
at a distance that follows $D(\alpha)$
and
after a waiting time $\tau$ given by 
the waiting-time probability density, $D(\tau)$. 
However, an important ingredient is missing in this picture,
as we need to take into account the correlations between
distances and waiting times.

With this goal in mind, we introduce the conditional 
waiting-time probability density, $D(\tau \, |\, X) $,
%%$$
%%D(\tau \, |\, X) \equiv \frac{\mbox{Prob}
%%[\tau \le \mbox{ waiting time }< \tau+d\tau \,|\, X]}{d\tau},
%%$$
where $|\, X$ means that only the cases where $X$ is verified
are taken into account; in our case $X$ will be a set of values of the
jumps.
It turns out that if $D(\tau \, |\, X)$ does not depend on $X$,
then $\tau$ and $X$ are independent,
whereas when $D(\tau \, |\, X)$ changes with $X$,
then $\tau$ and $X$ are correlated (nonlinearly, in general).

\begin{figure*}
\centering
\includegraphics[width=3.5in]{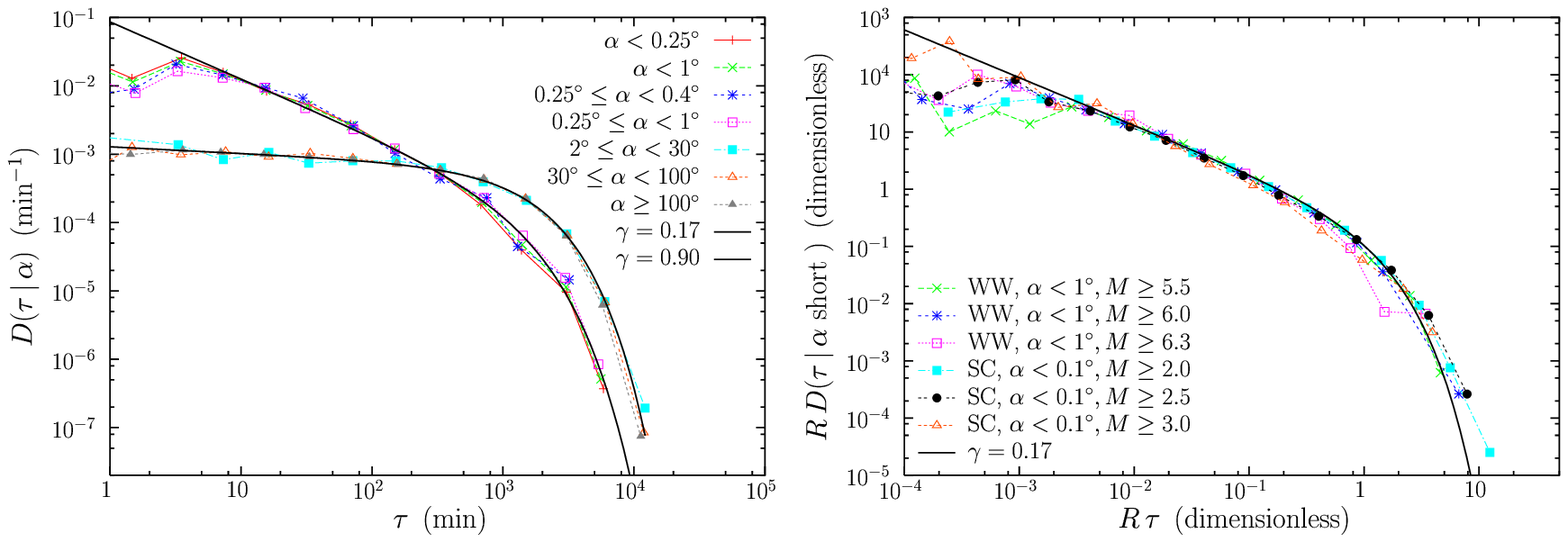}
\caption{(color online)
(a)
Waiting-time probability densities conditioned to different
sets of values of the jumps, in particular $\alpha$ short (below $1^\circ$)
and $\alpha$ long (above $2^\circ$), using worldwide seismicity 
with $M\ge 5.5$. In each case, gamma fits to all the curves are shown, 
with parameter $\gamma = 0.17$ ($\alpha$ short) and 
$\gamma=0.90$ ($\alpha$ long).
%%%%%%%%%%%%%%%%%%%%%%%%%%%%%%%%%%%%%%
(b)
Rescaled waiting-time probability densities conditioned to short 
jumps, $\alpha < 1^\circ$ for worldwide
seismicity (WW) and $\alpha < 0.1^\circ$ for Southern California
(SC).
Different $M_c$ values are used. 
The rescaling factor $R$ refers to the unconditional distribution,
and only depends on $M_c$ and on the spatial region $\mathcal{R}$.
The solid line is the gamma fit in (a) rescaled
in the same way.
\label{Dtconddist}
}
\end{figure*}

From the behavior of $D(\alpha)$, a natural threshold
for the jumps in the worldwide case is $\alpha \simeq 2^\circ$.
Figure \ref{Dtconddist}(a) compares $D(\tau \, |\, \alpha \mbox{ short })$
with $D(\tau \, |\, \alpha \mbox{ long })$ for worldwide
seismicity, where short and long refer to sets of distances
below and above $2^\circ$, respectively.
The differences are clear: for short jumps
the waiting-time distribution is a decreasing power law
for several decades, with an exponent close to 1,
and ends in an exponential decay.
We can identify these distributions with highly correlated events, 
i.e., aftershock sequences.
On the other hand, 
for long $\alpha$ the waiting-time distribution seems to be exponential
for its full range, compatible with a Poisson process
and therefore with independent occurrence.

But more surprising than the differences between short and long
jumps are perhaps the similarities for short jumps.
In fact, there seems to be a radical change of behavior
separated by $\alpha \simeq 2^\circ$, in the sense that if we
are above, or below, this value, the distributions do not change,
in other words, 
$D(\tau | \alpha < 0.25^\circ ) = D(\tau | 0.25 \le \alpha < 1^\circ )$,
etc. and 
$D(\tau | 2^\circ \le \alpha < 30^\circ )= D(\tau | \alpha \ge 100^\circ )$, 
etc., see Fig. \ref {Dtconddist}(a)
(for the range between $1^\circ$ and $2^\circ$ 
the behavior is not clear as the statistics is low).
This means that for short jumps the waiting-time distribution
is independent on the value of the jump, and the same happens
for long jumps, but when the whole range of jumps is considered
this is no longer true and both variables become dependent,
in contrast to Ref. \cite{Davidsen_distance}.
For each set of curves we can fit a gamma distribution, 
$D(\tau |\alpha) \propto e^{-\tau/a} / \tau^{1-\gamma}$,
turning out to be $\gamma \simeq 0.17$ for short $\alpha$
and $\gamma\simeq 0.9$ for long $\alpha$.
The latter value is very close to 1, the characteristic value
of a Poisson process; in fact, the difference between 1 and 0.9
is not significant within the uncertainty of our data,
but if the hypothesis of a Poisson process for long distances
could be rejected at any reasonable significance level 
(for which much more data would be necessary)
and a value $\gamma < 1$ could be significantly established,
this would constitute a support for the existence of long-range
earthquake triggering.
In any case, our findings are in disagreement with the hypothesis put forward 
in Ref. \cite{Huc_Main}, which considers seismicity as a process uncorrelated
in time but correlated in space.

This result confirms that the universal scaling law found 
for stationary seismicity  in Ref. \cite{Corral_prl.2004} 
is in fact a mixture of aftershocks and independent events,
turning out to be very striking that this mixture
leads to a universal behavior. 
The reason could be a universal proportion of aftershocks
versus mainshocks in stationary seismicity
\cite{Hainzl_bssa}.
%
%%Now that we have the distribution of jumps $D(\alpha)$
%%and the distribution of waiting times conditioned to the values
%%of $\alpha$, $D(\tau | X)$, we can use them as the input
%%of a continuous time random walk model to try to predict
%%the diffusion of earthquake occurrence.
%%(faltaria ver si las distancias pueden considerarse 
%%independientes unas de otras, en todo caso no creo 
%%que sea un efecto demasiado importante).
%
Further, the conditional time distributions
verify a scaling law, 
$$
D(\tau | \alpha_{\pm}) = R(\mathcal{R},M_c) f_{\pm} 
(R(\mathcal{R},M_c) \tau),
$$
where the indices $+$ and $-$ refer to long and short jumps,
respectively, and $R^{-1}(\mathcal{R},M_c)$ is the mean waiting time
of the unconditional distribution for the region $\mathcal{R}$
and $M\ge M_c$
(obviously, a scaling with the mean of the conditional distribution also
holds). Figure \ref{Dtconddist}(b) shows this behavior.

\begin{figure*}
\centering
\includegraphics[width=3.5in]{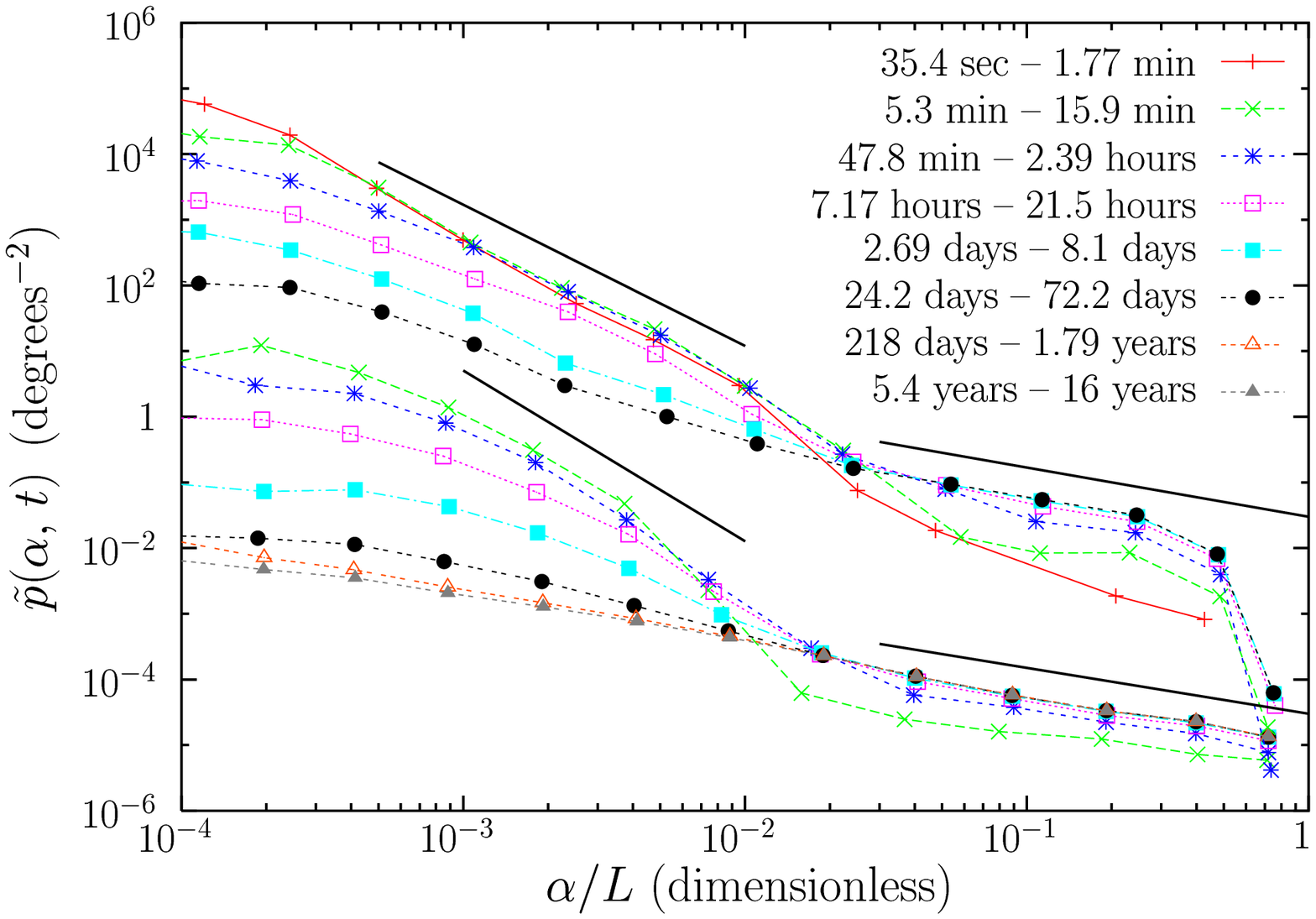}
\caption{(color online)
%%%Distribution of earthquake distances after a time $t$,
Diffusion profile $\tilde p(\alpha,t)$ versus $\alpha$
(rescaled by $L$)
for worldwide seismicity with $ M \ge 5.5$ (bottom curves)
and for Southern California with $ M \ge 2.5$ (top).
Time ranges from tens of seconds to several years.
The displayed power laws are indicative, 
and have the same exponents as the jump distributions.
\label{p}
}
\end{figure*}

The next step is to measure earthquake diffusion 
directly from data.
The fundamental quantity is $p(\alpha,t)$,
which we call diffusion profile and
gives the probability density that two earthquakes
(not consecutive) are at a distance $\alpha$ when
they are separated by a time $t$, i.e.,
$$
p(\alpha,t) d\alpha \equiv {\mbox{Prob}
[\alpha \le \mbox{ distance at a time } t < \alpha+d\alpha}].
$$
In practice, the single value of $t$ is replaced by a range of values
and, as above, we introduce 
$\tilde p(\alpha,t) \equiv p(\alpha,t) / (2 \pi \sin \alpha)$.
In the case of normal diffusion $\tilde p(\alpha,t)$ is given by a Gaussian
distribution (more precisely, a semi-Gaussian, as $\alpha > 0$),
with a second moment scaling as $\langle \alpha^2 \rangle \sim t$,
whereas $p(\alpha,t)$ is given by a Rayleigh distribution
(or a Maxwell distribution in three dimensions).
In contrast, our measurements 
yield to results far from normality, see Fig. \ref{p}.
For short times, the profile resembles the distribution of jumps
with 2 regimes; as time evolves events migrate farther 
from the origin, towards the long-range part of the 
curve, which becomes dominant for all $\alpha$ 
for long times, with an exponent of value 0.8 for $\tilde p$.
%%%%(for $M\ge 5.5, WW)$.

A scaling law also holds for the diffusion profile, i.e.,
$$
%p(\alpha,t) = \frac 1 L h\left(\frac \alpha L, t \right).
p(\alpha,t) = h(\alpha/ L, t)/L.
$$

Finally, the process turns out to be clearly subdiffusive,
as $\langle \alpha^2 \rangle $ grows more slowly than linearly,
although we cannot provide a definite value of a exponent
as in Refs. \cite{Marsan00,Helmstetter_jgr03}.

This paper would not have been possible
without the effort of the researchers who have
made earthquake catalogs publicly available.
The Ram\'on y Cajal program and the projects
BFM2003-06033  (MCyT) and
2001-SGR-00186 (DGRGC) are acknowledged.

%%%\end{document}

%%\newpage

%%\bibliographystyle{apalike}
%\bibliographystyle{unsrt}

%\bibliography{../../biblio}

%\newpage

\end{document}